\documentclass[a4paper,11pt]{article}
\pdfoutput=1

\usepackage{jinstpub}

\usepackage{graphicx}
\usepackage[version=4]{mhchem}
\usepackage{float}
\usepackage{subfigure}

\usepackage{url}
\usepackage{makecell}
\usepackage{siunitx}

\usepackage[utf8]{inputenc}

\title{Detector control system for the CBM-TOF}

\author[a,1]{S. Dong}
\author[a,1]{G.M. Huang\note{Corresponding author.}}
\author[b]{J. Fr\"uhauf}
\author[b]{P.-A. Loizeau}
\author[c]{I. Deppner}
\author[c]{N. Herrmann}
\author[a]{D. Wang}

\affiliation[a]{Central China Normal University,\\No.152 Luoyu Road, 430079 Wuhan, China}
\affiliation[b]{GSI Helmholtzzentrum f\"ur Schwerionenforschung GmbH,\\Planckstra{\ss}e 1, 64291 Darmstadt, Germany}
\affiliation[c]{Physikalisches Institut der Universit\"at Heidelberg,\\Im Neuenheimer Feld 226, 69120 Heidelberg, Germany}

\emailAdd{s.dong@mails.ccnu.edu.cn}
\emailAdd{gmhuang@mail.ccnu.edu.cn}

\abstract{A high-performance time-of-flight (TOF) MRPC wall is being built for the CBM experiment at FAIR for charged hadron identification. The detector control system for the TOF system will be based on EPICS. All components like power supplies for low and high voltages, power distribution boxes, gas control and front-end electronics (FEE) are controlled and monitored. In a test, called mini-CBM, all these functionalities are implemented and tested. For monitoring the detector environment and the status of the front-end electronics, a slow control application is implemented based on IPbus, which is an FPGA-based slow control bus used for the TOF data acquisition system. In addition to the functions of control and monitoring, exception handling and data archiving services are implemented as well. This system has been fully verified in beam tests in 2019 at GSI.
}


\proceeding{XV Workshop on Resistive Plate Chambers and Related Detectors - RPC2020\\
10-14 February, 2020 \\
Rome, Italy}

\begin{document}
\maketitle
\flushbottom

\section{Introduction}
% Intro CBM
The Compressed Baryonic Matter~(CBM) experiment is a heavy-ion experiment located at the Facility for Antiproton and Ion Research~(FAIR) in Darmstadt, Germany~\cite{ablyazimov2017challenges}. The experiment will use high-energy nucleus-nucleus collisions to study the equation-of-state of nuclear matter at neutron star core densities, and search for phase transitions, chiral symmetry restoration, and exotic forms of QCD matter.

% Intro TOF
A \SI{120}{\square\meter} large time-of-flight (TOF) wall composed of multigap resistive plate chambers~(MRPCs) is one key element in the CBM~\cite{herrmann2014technical} experiment. The TOF wall provides hadron identification at incident beam energies from \SI{2}{\ampere \GeV} to \SI{11}{\ampere \GeV}. The system time resolution should be better than \SI{80}{\ps}, including all electronic jitter and the resolution of the time reference system. The reaction rate can reach up to \SI{10}{\MHz} to achieve the required statistics and precision. The MRPCs will be exposed to a particle flux between \SI{0.1}{\kHz/\cm\square} and \SI{100}{\kHz/\cm\square} depending on their location. The TOF wall's actual conceptual design is composed of 6 different types of modules. There will be 226 modules in total.

% Intro mCBM
The mini-CBM (mCBM) test-setup is an important project towards the realization of the CBM experiment~\cite{cbm2017mcbm}. This setup is installed at the SIS18 facility of GSI/FAIR. It includes detector modules from all CBM subsystems using prototypes or (pre-)series production modules. The setup will allow us to test and optimize the detector performance as well as the firmware and software chain under realistic experimental conditions. This will reduce significantly the commissioning time for the CBM experiment. A total of 5 real size CBM Prototype modules (25 MRPCs) are installed in the mCBM test setup.

% Details of design 
\section{The Detector Control System of the mCBM}
% Motivation
The Experimental Physics and Industrial Control System (EPICS)~\cite{dalesio1994experimental} is widely used in control systems to operate devices such as particle accelerators, large experiments, and major space telescopes. It is well suited to monitor and control experimental equipment like low voltage and high voltage power supplies, temperature sensors and pressure gauges. It provides specific network protocols, such as channel access and pvAccess (process variable access), for high-bandwidth soft real-time networking applications. Due to these advantages, EPICS has been chosen to build the detector control system (DCS) of the CBM experiment.

The instruments and devices controlled by the mCBM test stand are listed in Table~\ref{tbl:Instruments}. A total of 2635 process variable (PV) channels are created and archived resulting in a total of approximately 200 MB of disk storage per day of operation. The mCBM as a prototype test stand needs full functionality like the CBM just with reduced endpoints. All features like archiving, safety aspect, online monitoring and stability tests can be proven at this point. Scalability is of course an issue as well since it is needed to increase the system without rewriting everything from scratch.

\begin{table}[h]
   \centering
   \begin{tabular}{ l l l l }
      \hline
      \textbf{Instrument}     & \textbf{Model}                &\textbf{Quantity}               & \textbf{Protocol} \\
      \hline
      Low voltage             & \makecell[l]{MeanWell RCP-2000-12\\MeanWell RKP-1UI-CMU1}  &  \makecell[l]{1\\1}  &HTTP \\
      \hline
      \makecell[l]{Low-voltage distributor\\ Low-voltage channels} & GSI Low-voltage distributor                 &  \makecell[l]{1\\16}  &\makecell[l]{RS232\\I2C}         \\
      \hline
      High voltage            & \makecell[l]{CAEN SY1527LC\\ CAEN A1526P \\ CAEN A1526N} & \makecell[l]{1\\3\\3} &TCP/IP \\
      \hline
      Gas flow                & Bronkhorst control units      & 2 & I2C           \\
      \hline
      FEE                     & TOF front-end electronics                 & 10      & IPbus         \\
      \hline
      Detector environment    & TOF front-end electronics              & 10         & IPbus         \\
      \hline
      Readout channels    & TOF front-end electronics              & 1600         & IPbus       \\
      \hline
   \end{tabular}
   \caption{Instruments that are controlled and monitored in the mCBM}
   \label{tbl:Instruments}
\end{table}

% Intro EPICS
An EPICS control application typically contains three parts:
\begin{itemize}
   \item Operator interfaces (OPIs), 
   \item Input-Output controllers (IOCs)
   \item Local area network (LAN).
\end{itemize}

\begin{figure}[ht]
   \centering
   \includegraphics[width=.9\textwidth]{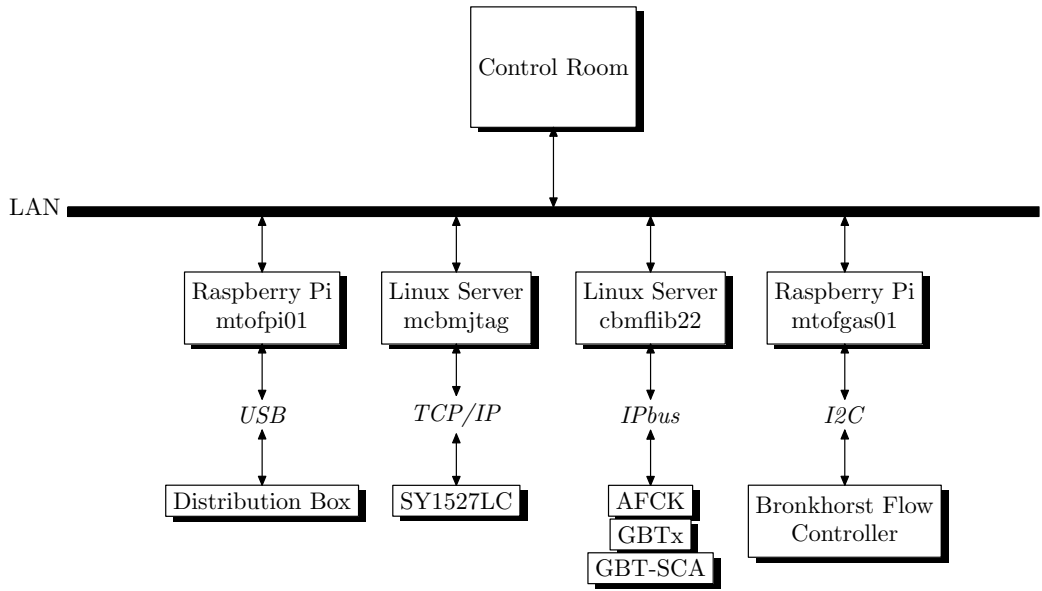}
   \caption{The TOF DCS physical structure.}
   \label{fig:TOF DCS structure}
\end{figure}

An OPI is a workstation that can run various EPICS tools such as the Motif Editor and Display Manager (MEDM) and Control System Studio (CSS)~\cite{hatje2007control}. An IOC supports the EPICS run-time databases together with the other software components such as device support, device drivers, sequencer and database access. The LAN is the communication network that allows the IOCs and OPIs to communicate. The physical structure of the CBM-TOF DCS is shown in Figure~\ref{fig:TOF DCS structure}. The OPIs of the mTOF DCS are designed using CSS. The IOCs can communicate with the hardware through various interfaces, such as USB, I2C, and Ethernet. 

% Intro Raspberry Pi
A Raspberry Pi~\cite{RaspberryPi:online} is a small single-board computer that contains several interfaces. EPICS is well supported by Raspberry Pi OS (formerly Raspbian), which is a Debian-based operating system. Improvements in the hardware and overall performance allow this small and inexpensive platform to run various instances of IOCs to control several devices and instruments.

% Power Supplies
\subsection{Power Supplies and Gas Flow}
The HV and LV power supply systems for the TOF wall operation are placed within the CBM cave. They will be protected from excessive radiation by concrete blocks. All power supplies are controlled via the EPICS system.

\begin{figure}[ht]
   \begin{center}
      \includegraphics[width=0.8\textwidth]{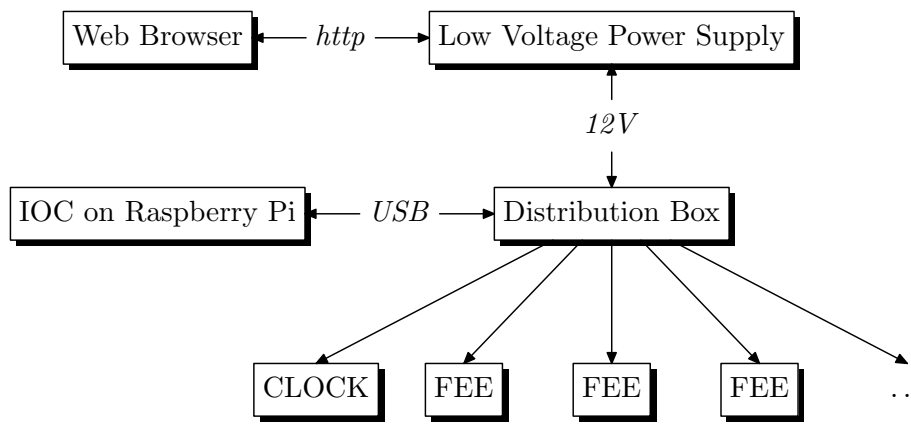}
      \caption{The structure of the mTOF low-voltage power distribution system.}
      \label{fig:mTOF LV structure}
   \end{center}
\end{figure}

% Low voltage
A low-voltage power distribution system is designed to provide power for the FEE and the clock system. The LV system consists of three devices, a commercial power supply, a distribution box and a Raspberry Pi. The power consumption of each LV channel is approximately \SI{60}{\W}. Therefore, a power fan-out system was developed that can handle high input current from the power supply and support multiple channels controllable via the EPICS system. Figure~\ref{fig:mTOF LV structure} depicts the structure of the low-voltage power supply used in the mCBM.

% Introduce devices
A MeanWell power supply, containing a control/monitor unit (RKP-1UI-CMU1) and one rack power supply (RCP-2000-12), is used. The power supply provides \SI{12}{\volt} with up to \SI{100}{\ampere} on its output. The power distribution box  splits the provided \SI{12}{\volt} from 1 input channel to 16 output channels. In total 6 channels are required for the mTOF system, 5 channels for the FEEs, and one for the clock system. Other channels are used for test counters and spares for future upgrades of the mTOF system.

% Introduce control
The MeanWell power supply as well as the distribution box is remote controlled. Up to now, the Power Supply is controlled via the provided user interface accessible via Web Browser. The distribution box is controlled by a Raspberry Pi.

The primary device in the distribution box is an Arduino~\cite{Arduino:online} board. It controls the output status of each low voltage channel and reads out the current through I2C devices. The connection between the Arduino and the Raspberry Pi is made via USB. String commands are used to control the Arduino. The EPICS StreamDevice~\cite{StreamDevice:online} module is chosen to build the IOC. This module provides a generic EPICS device support for devices with a "byte stream"-based communication interface. The slave devices can be controlled by sending and receiving strings. It is possible to switch on/off each channel and obtain the current information by sending the string commands from the Raspberry Pi to the Arduino board.

\begin{figure}[ht]
   \begin{center}
     \includegraphics[width=.8\textwidth]{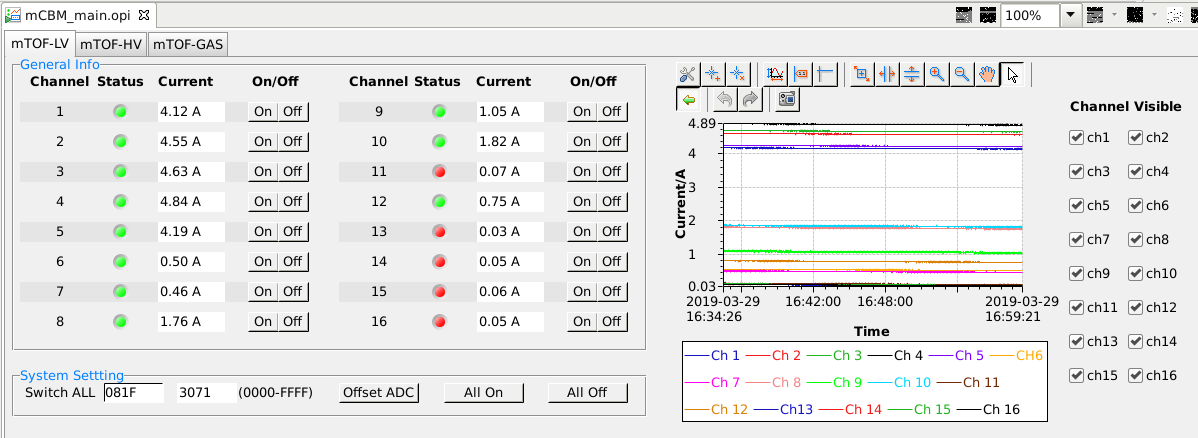}
     \caption{A snapshot of the graphical user interface of the mTOF low-voltage distributor.}
     \label{fig:LV OPI}
   \end{center}
\end{figure}

% High voltage
Besides the LV, each MRPC Module has two HV inputs, one positive and one negative HV channel. The mTOF system requires $2\times5$ HV channels plus some additional channels for test counters. The final design of the TOF wall will need more than $2\times200$ HV channels. Up to now a CAEN SY1527LC crate with three pairs of HV modules (CAEN A1526P/N) is used to power the MRPC modules. All operational parameters of the HV system are controlled and monitored via the built-in links (RS232, H.S. CAENET, Ethernet). We use the Ethernet interface to control and monitor the HV systems remotely from the control room.

The EPICS IOC of HV is designed using the AsynDriver~\cite{kraimer2005epics} module. This EPICS module is a general-purpose facility for interfacing device-specific code to low-level drivers. The IOC connects via Ethernet to the crate and then controls and monitors the HV modules parameters through the driver.

% GAS flow
The gas supply system for the CBM-TOF wall consists of the following units: gas supplies, a gas mixer, a distribution manifold, active volume counters, a collection manifold, gas purification, a gas analysis/monitoring unit and recycling exhaust. The system is built as a closed-loop system. The MRPC detector’s operation relies on reasonable control of the working gas and monitoring of the environmental conditions.

At the mCBM a gas control unit, designed by the Laboratory of Instrumentation and Experimental Particle Physics (LIP), is used. This unit can control the working gases (\ce{i-C4H10}, \ce{C2H2F4}, \ce{SF6}) of the MRPC by several Bronkhorst control units in the crate. These devices control the flow rate of each gas separately. The communication unit inside the crate is a Raspberry Pi. The Bronkhorst control units are connected via the I2C protocol. The Bronkhorst units can be controlled and monitored with an EPICs IOC by using the AsynDriver module. Furthermore, the gas parameters like pressure, relative humidity, and the environmental parameters (temperature, atmospheric pressure) are read out as well. 

% FEE
\subsection{FEE and Detector Environment}
\begin{figure}[ht]
	\begin{center}
		\includegraphics[width=1.0\textwidth]{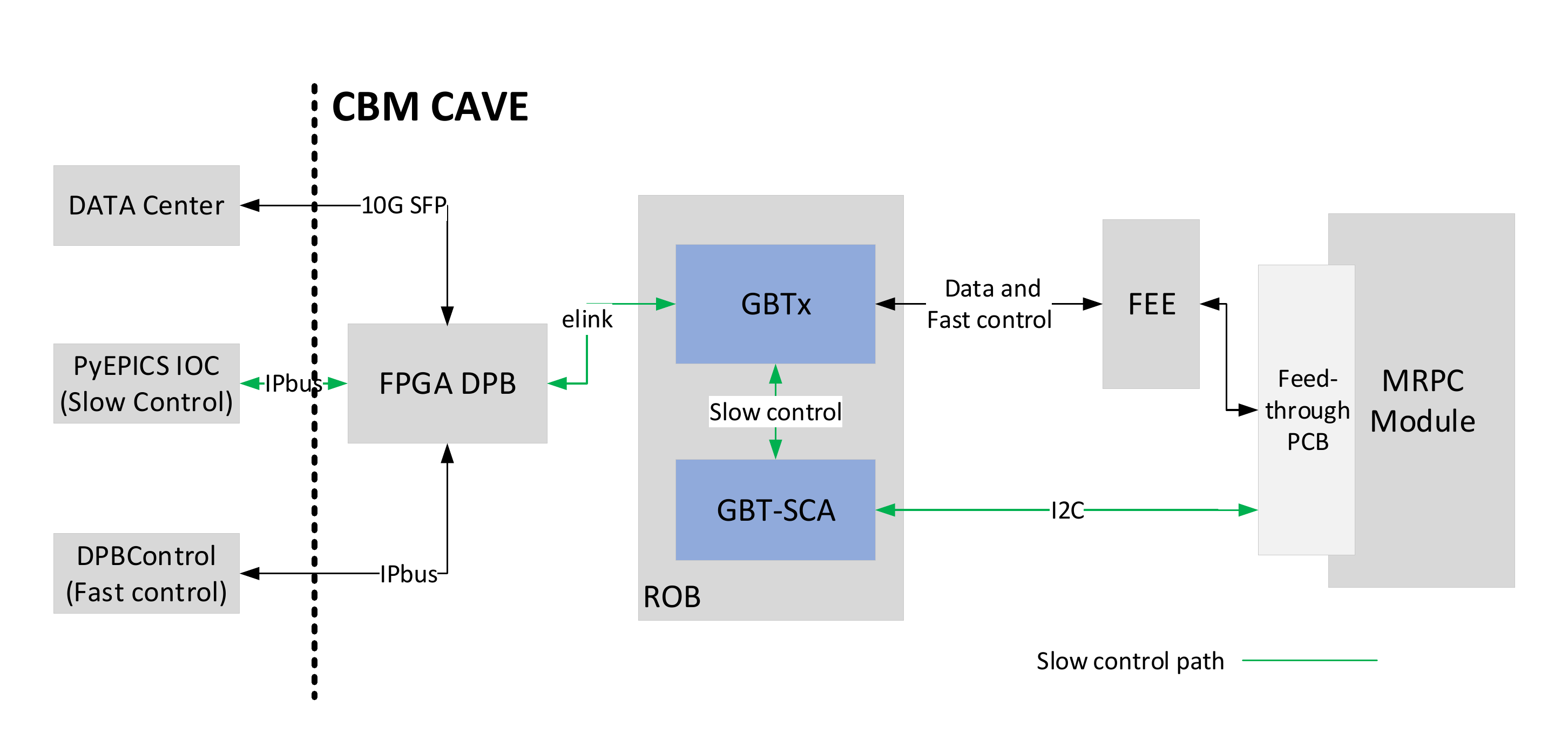} 
	\caption{The structure of the fast control link and the slow control link of the mTOF.}
	\label{fig:readout-chain}
	\end{center}
\end{figure}

The readout electronic of the mTOF is placed close to the detector in the experiment cave. Radiation hardness therefore plays an important role in its design. The gigabit transceiver (GBT) project~\cite{Moreira:2009pem} devices are designed to interface with on-detector and off-detector electronics in a radiation environment. The GBTx transceiver ASIC and GBT-SCA~\cite{caratelli2015gbt} ASIC are equipped on the readout board (ROB) to interconnect the FEE inside the cave with the FPGA based data processing board (DPB) in the DAQ Room with an optical link. To communicate with the DPB the IPbus ~\cite{twepp14_ipbus} protocol is used. Figure~\ref{fig:readout-chain} presents the structure of the readout chain of the mTOF. The GBT-SCA ASIC builds the interface for control and monitoring signals on the FEE. It provides several user I/Os, DACs, ADCs as well as I2C Master and SPI Master interfaces.

To monitor the detector environment, one sensor (BME280) is equipped on each feed-through PCB on the inside of the GAS Box. It is connected to one of the I2C Ports of the GBT-SCA sitting on the ROB. The read-out and histogrammed parameters of this sensor are humidity, temperature, and pressure. Besides the I2C ports, ADC ports are used to monitor the different voltages on the FEE boards. The temperature of the GBT-SCA is read out and with one GPIO port of the GBT-SCA a switch is controlled to choose between the GBTx recovered clock and an external \SI{160}{\MHz} clock for the FEE. Four LEDs are connected to the GPIO Pins to indicate whether the PCB is working as expected.

\begin{figure}[ht]
	\begin{center}
		\includegraphics[width=1.0\textwidth]{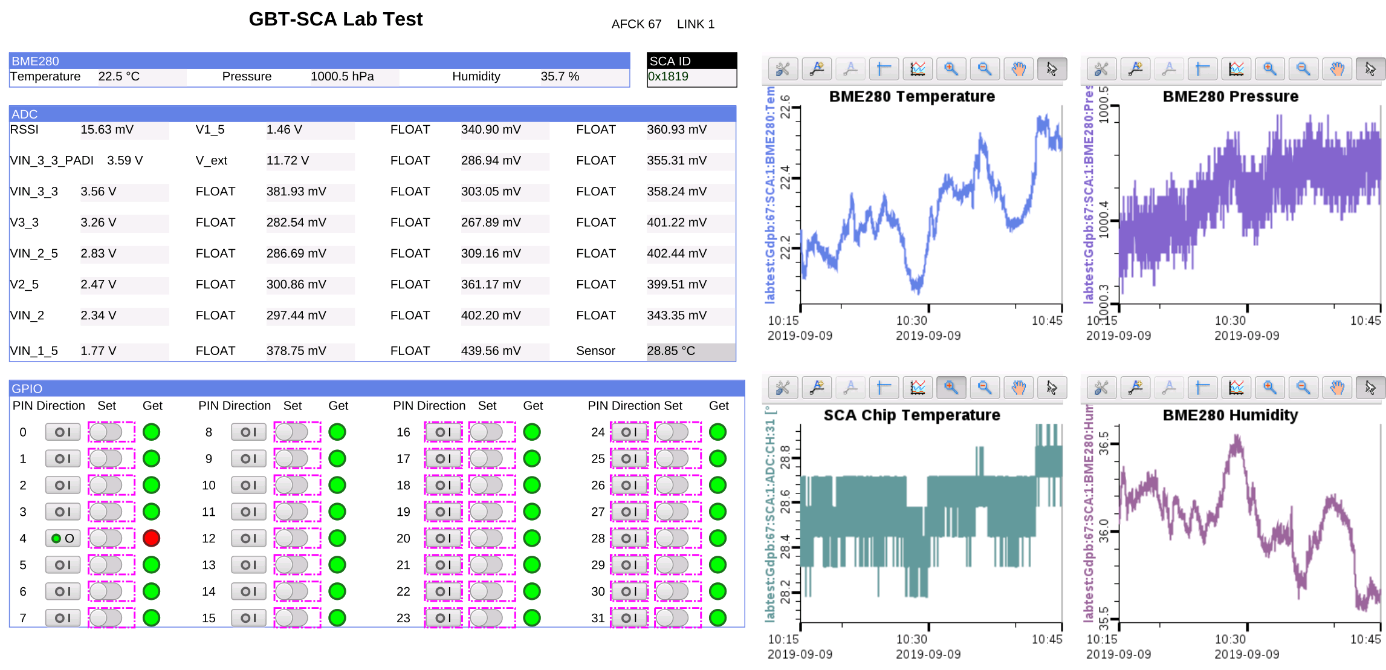} 
	\caption{An example of the SCA user interface }
	\label{fig:opi-sca-readout}
	\end{center}
\end{figure}

The data from the GBT-SCA are passed via the GBTx communication path and transferred along with the detector and FEE data to the DPB. To control and monitor the data from the GBT-SCA ASIC within the EPICs environment software, written in Python, an IPbus Module inside the DPB firmware, written in VHDL, was developed. Figure~\ref{fig:opi-sca-readout} show an example of the SCA user interface as well as the data provided by one sensor.

\subsection{Data Archiving and Exception Handling}
% Archiving
The collected slow control information is an important tool to verify the detector status. For example, the HV current is an indicator for hit rates on the one hand and provides information of the detector healthiness on the other hand. Environmental data like temperature data can be helpful later on if there is a non-understood drift inside the Detector data for example. Therefore, all slow control data are archived into a database during beam tests.

\begin{figure}[ht]
   \centering
   \begin{minipage}{.5\textwidth}
      \centering
      \includegraphics[width=.90\textwidth]{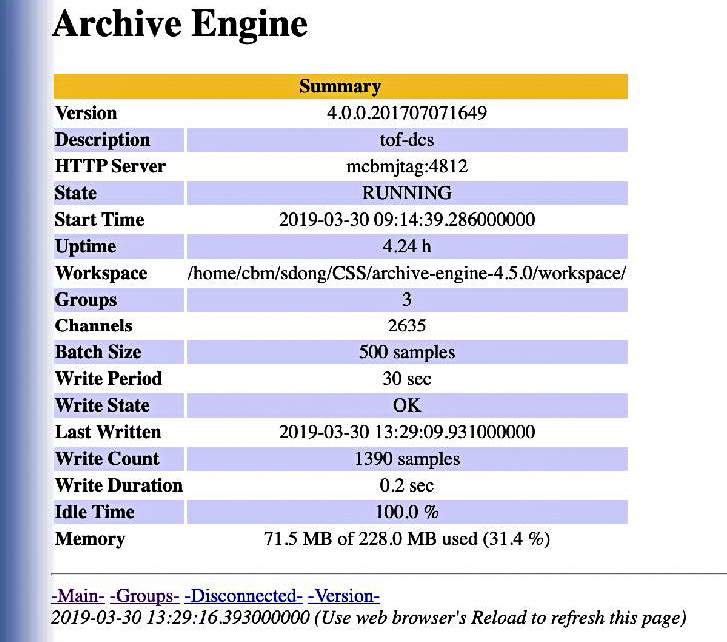} 
      \caption{The status of archive engine.}
      \label{fig:archive-status}
   \end{minipage}%
   \begin{minipage}{.5\textwidth}
      \centering
      \includegraphics[width=.96\textwidth]{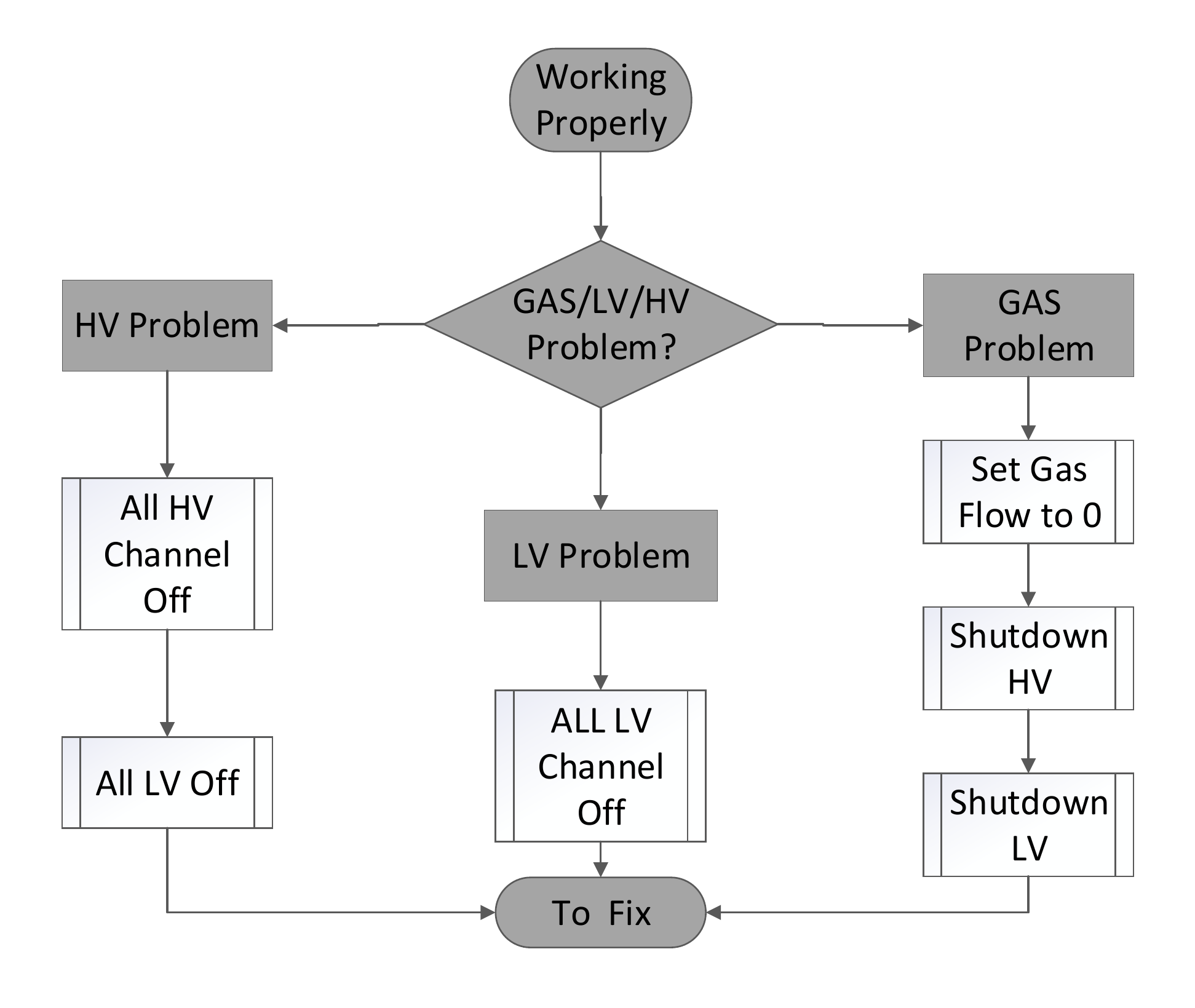} 
      \caption{The exception handling flowchart for mTOF.}
      \label{fig:exception-handle}
   \end{minipage}
\end{figure}

As one part of CSS, the archive system~\cite{ArchiveSystem} is applied to archive the slow control data of the mTOF system. The archive engine of this system takes the slow control data from EPICS IOCs via channel access and pushes them into the database. The period of the PV data reading is 1 second. It is possible to use the archive tools to view the archive system’s status by a web browser. This opens the opportunity to check the system’s condition at any time from outside of the experiment cave. Figure~\ref{fig:archive-status} shows the archive engine status at the mCBM.

The mTOF slow control system database is designed using PostgreSQL~\cite{PostgreSQL}, an open-source object-relational database system with a strong reputation for reliability and performance. 

% Exception handling
The DCS needs to be very robust and stable since any failure in the HV or Gas system can harm the detector. Controlled action in terms of safe operation in case of an error condition occurs needs to be implemented. The Sequencer~\cite{lupton2000state} EPICS module is used to implement such state transition. Figure~\ref{fig:exception-handle} shows that the exception handling which is implemented using this module. For example, if some problems occur on an HV or gas system, all HV channels should be shut down to protect the detector. The LV system will ramp down once the HV is fully turned off.

\section{Conclusions}
\label{sec:Conclusions}
The DCS for the CBM-TOF is designed using EPICS and has been operated and validated during beam tests at GSI. Besides the hardware interfaces for the detector components a control path for the data provided by the FEE and DAQ system is developed. 

\bibliographystyle{JHEP}
\bibliography{DetectorControlSystemforCBMTOF}

\providecommand{\href}[2]{#2}\begingroup\raggedright\begin{thebibliography}{10}

\bibitem{ablyazimov2017challenges}
T.~Ablyazimov, A.~Abuhoza, R.~Adak, M.~Adamczyk, K.~Agarwal, M.~Aggarwal
  et~al., \emph{Challenges in qcd matter physics--the scientific programme of
  the compressed baryonic matter experiment at fair}, {\emph{The European
  Physical Journal A} {\bfseries 53} (2017) 1}.

\bibitem{herrmann2014technical}
N.~Herrmann et~al., \emph{Technical design report for the cbm time-of-flight
  system (tof)}, {\emph{GSI, Darmstadt} (2014) }.

\bibitem{cbm2017mcbm}
\url{https://fair-center.eu/en/for-users/experiments/nuclear-matter-physics/cbm/projects/mcbm.html}.

\bibitem{dalesio1994experimental}
L.~R. Dalesio, J.~O. Hill, M.~Kraimer, S.~Lewis, D.~Murray, S.~Hunt et~al.,
  \emph{The experimental physics and industrial control system architecture:
  past, present, and future}, {\emph{Nuclear Instruments and Methods in Physics
  Research Section A: Accelerators, Spectrometers, Detectors and Associated
  Equipment} {\bfseries 352} (1994) 179}.

\bibitem{hatje2007control}
J.~Hatje, M.~Clausen, C.~Gerke, M.~Moeller and H.~Rickens, \emph{Control system
  studio (css)}, {\emph{ICALEPCS07, Knoxville, TN, USA} (2007) }.

\bibitem{RaspberryPi:online}
\url{https://www.raspberrypi.org/products/raspberry-pi-3-model-b/}.

\bibitem{Arduino:online}
\url{https://www.arduino.cc/}.

\bibitem{StreamDevice:online}
D.~Zimoch, ``Epics streamdevice.''
  \url{http://epics.web.psi.ch/software/streamdevice/}, 2017.

\bibitem{kraimer2005epics}
M.~R. Kraimer, M.~Rivers and E.~Norum, \emph{Epics: Asynchronous driver
  support},  in \emph{Proc. Int. Conf. Accelerator and Large Experimental
  Physics Control Systems}, pp.~074--5, 2005.

\bibitem{Moreira:2009pem}
P.~Moreira et~al., \emph{{The GBT Project}},  in \emph{{Topical Workshop on
  Electronics for Particle Physics}}, 2009,
  \href{https://doi.org/10.5170/CERN-2009-006.342}{DOI}.

\bibitem{caratelli2015gbt}
A.~Caratelli, S.~Bonacini, K.~Kloukinas, A.~Marchioro, P.~Moreira,
  R.~De~Oliveira et~al., \emph{The gbt-sca, a radiation tolerant asic for
  detector control and monitoring applications in hep experiments},
  {\emph{Journal of Instrumentation} {\bfseries 10} (2015) C03034}.

\bibitem{twepp14_ipbus}
C.~Ghabrous~Larrea, K.~Harder, D.~Newbold, D.~Sankey, A.~Rose, A.~Thea et~al.,
  \emph{{IPbus: a flexible Ethernet-based control system for xTCA hardware}},
  \href{https://doi.org/10.1088/1748-0221/10/02/C02019}{\emph{JINST} {\bfseries
  10} (2015) C02019}.

\bibitem{ArchiveSystem}
J.~Hatje, M.~Clausen, C.~Gerke, M.~Moeller and H.~Rickens, ``Cs-studio guide,
  chapter 11. archive system.''
  \url{http://cs-studio.sourceforge.net/docbook/ch11.html}, 2019.

\bibitem{PostgreSQL}
T.~P. G.~D. Group, ``Documentation postgresql 10.13.''
  \url{https://www.postgresql.org/docs/10/index.html}, 2019.

\bibitem{lupton2000state}
W.~Lupton and B.~Franksen, ``State notation language and sequencer users\rq{}
  guide.''
  \url{https://www.slac.stanford.edu/grp/ssrl/spear/epics/site/seq/Manual.pdf},
  2000.

\end{thebibliography}\endgroup

\end{document}